\documentstyle[preprint,version2,aps,epsf,rotate]{revtex}

\begin{document}
\draft
\begin{title}
\begin{center}
The effect of magnetic dipolar interactions on \\ the interchain spin
wave dispersion in CsNiF$_3$
\end{center}
\end{title}

\author{M. Baehr, M. Winkelmann, P. Vorderwisch, M. Steiner\\
$^\ast$C. Pich, $^\ast$F. Schwabl}
\begin{instit}
BENSC, Hahn-Meitner-Institut, Glienickerstr. 100, 14109 Berlin, Germany \\

$^\ast$TU-M\"unchen, James-Franck-Strasse, 85747 Garching, Germany
\end{instit}
                                    
\begin{abstract} Inelastic neutron scattering measurements were 
  performed on the ferromagnetic chain system CsNiF$_3$ in the
  collinear antiferromagnetic ordered state below $T_N = 2.67$~K.  The
  measured spin wave dispersion was found to be in good agreement with
  linear spin wave theory including dipolar interactions. The
  additional dipole tensor in the Hamiltonian was essential to explain
  some striking phenomena in the measured spin wave spectrum: a
  peculiar feature of the dispersion relation is a jump at the zone
  center, caused by strong dipolar interactions in this system.  The
  interchain exchange coupling constant and the planar anisotropy
  energy were determined within the present model to be $J'/k_B =
  -0.0247(12)$~K and $A/k_B = 3.3(1)$~K. This gives a ratio $J/J'
  \approx 500$, using the previously determined intrachain coupling
  constant $J/k_B = 11.8$~K.  The small exchange energy $J'$ is of the
  same order as the dipolar energy, which implies a strong competition
  between the both interactions.
\end{abstract}

 \narrowtext

\section{Introduction}

The compound CsNiF$_3$ is the best known example of a quasi
one-dimensional (1D) ferromagnet. It crystallizes in the hexagonal
ABX$_3$ type structure (P6$_{3}$/mmc, with $a = b =6.21${\AA} and $c =
5.2$\AA) \cite{BABE69}\@.  The Ni$^{2+}$ ions (S=1) are located in the
centers of NiF$_6$-octahedra, which are linked by common faces to form
chains along the $c$ axis.  A series of fundamental investigations on
linear and non-linear spin dynamics above $T_{\rm N}$ \cite{STEI74}\@ and in an
external magnetic field have been performed \cite{STEI76,STEI91}\@.
Below $T_N=2.67$~K three-dimensional ordering sets in due to an isotropic
interchain interaction and the dipolar interaction. The Hamiltonian describing
the three-dimensional magnetic properties of CsNiF$_3$ is
\begin{equation}
  H = -2J\sum_{i}{\bf S}_i{\bf S}_{i+1}+A\sum_{l} {(S_l^z)}^2 -
  \sum_{l,l'}\left(J_{ll'}'\delta^{\alpha\beta}+ 
    A^{\alpha\beta}_{ll'}\right)S_l^\alpha S_{l'}^\beta \ .
  \label{hamilton}
\end{equation}
The index $i$ indicates positions on single spin chains, whereas
$l$ indicates all spin positions. In Eq. (1) the first two terms are
responsible for the one-dimensional behavior, i.e. $J$ denotes the
ferromagnetic intrachain interaction and $A$ the single-ion anisotropy. The
last two terms lead to the three-dimensional order, where 
$J_{ll'}'$ denotes the nearest neighbor interchain interaction and
$A^{\alpha\beta}_{ll'}$ the long-range dipolar interaction
\begin{equation}
  A^{\alpha\beta}_{l,l'} = -{(g\mu_B)^2\over
    2}\left\{{\delta^{\alpha\beta}\over |{\bf x}_l -{\bf
      x}_{l'}|^3}+{3({\bf x}_l -{\bf x}_{l'})^\alpha ({\bf x}_l -{\bf
      x}_{l'})^\beta \over |{\bf x}_l -{\bf x}_{l'}|^5}\right\}
\label{Dipoltensor}
\end{equation}
The coupling constant along the chain $J$ and the anisotropy 
energy $A$ were determined by inelastic neutron scattering in the one
dimensional ordered state (T $>$ T$_{\rm N}$) to be $J/k_B = 11.8$~K
and $A/k_B = 4.5$~K \cite{STEI76}\@. These values are based on linear
spin wave theory for classical spins \cite{VILL73}, whereas a larger
anisotropy constant $A = 9.0 K$ was determined, using a renormalized
spin wave theory for $S = 1$ spins \cite{LIND76} considering the
continuous degeneracy of the ground state. In both analyses the third and
fourth term of Eq. (\ref{hamilton}) had been neglected, which are important for
the three-dimensional ordering of CsNiF$_3$ especially the dipolar interaction
as indicated by the antiferromagnetic, collinear ordered ground state
\cite{STEI74}. 

In the three-dimensional state a purely isotropic antiferromagnetic exchange
coupling leads to a frustrated 120$^{\circ}$ structure in hexagonal ABX$_3$
compounds \cite{GAUL89,KADO87,YELO73}\@. In the limit of pure dipole
interaction a ferromagnetic spin arrangement is favored as in the case of a
pure two-dimensional triangular lattice \cite{Pich94,pich96}. However, if
dipolar and exchange energies are of the same order a collinear
antiferromagnetic structure occurs which will be shown later.
Due to the collinear order of the spins, the ground state is no more
continuously degenerate but shows three different domains (A,B,C), as
shown in Fig.~(\ref{Spinstruct0})~\cite{STEI72,STEI74}\@.

While the spin dynamics above $T_N$ are well known, the spin wave
excitations in the ordered state ($T<T_N$) have not yet been
studied in detail. The aim of the present investigation is to
determine the interchain coupling constant and to probe the effects of
the dipolar interaction on the spin wave spectrum. The evaluation
of the interchain magnon dispersion relation and the related neutron
scattering cross sections were performed, using quantum mechanical
spin wave theory including long-range dipolar interactions.

\section{Theory}
In this chapter we derive the excitation spectrum and the scattering amplitudes
within the linear spin-wave theory for the Heisenberg Hamiltonian in
Eq. (\ref{hamilton}). 

\subsection{Excitation spectrum}

In this section the dispersion relation for domain A will be derived.  In the
following we choose the Cartesian coordinate system shown in
Fig.~(\ref{Spinstruct0}) and the Brillouin zone as in Fig.~(\ref{BrillFig})\@.
Fourier transform of the Hamiltonian (Eq.~(\ref{hamilton})) yields
\begin{equation}
  H = -\sum_{\bf q} \left( J_{\bf q}\delta^{\alpha\beta}-
    A\delta^{\alpha z}\delta^{\beta z}+J_{\bf q}'\delta^{\alpha\beta}+
    A^{\alpha\beta}_{\bf q}\right)S_{\bf q}^\alpha S_{-{\bf q}}^\beta \ 
  ,
\label{HH}
\end{equation}
with the nearest-neighbor exchange energies (intrachain and interchain)
\begin{eqnarray}
  J_{\bf q} & = & 2J\cos q_z \\ J_{\bf q}'& = & 2J'\left(\cos q_x
    +2\cos{q_x\over 2}\cos{\sqrt 3q_y\over 2}\right) \label{Jstrichq} \, .
\end{eqnarray}
The Fourier transform of the dipole tensor
$A^{\alpha\beta}_{\bf q}$, is obtained via Ewald summation
technique\cite{Bonsal77}\@. The Holstein-Primakoff transformation, which
introduces Bose operators $a_l$ and 
$\ a_l^{\dag} $, is given up to bilinear order\cite{Ziman69} by
\begin{mathletters}
\begin{eqnarray}
  S_l^x  &=&  \pm(S-a_l^\dagger a_l), \label{trafo1}\\
  S_l^y &=&  \sqrt{S\over 2}(a_l+a_l^\dagger),\label{trafo2}\\
  S_l^z  &=&  \mp i\sqrt{S\over 2}(a_l-a_l^\dagger)\label{trafo3},
\end{eqnarray}
\end{mathletters}
where the upper (lower) sign corresponds to the first (second)
sublattice.  After Fourier transformation of these equations and
insertion into the Hamiltonian (Eq.~(\ref{HH})), regarding only
wave vectors perpendicular to the chain axis ($q_z=0$), the bilinear
term becomes
\begin{eqnarray}
  H^{(2)}  &= & \sum_{\bf q}A_{\bf q}~a_{\bf
    q}^{\dagger}a_{\bf q} + {1\over 2} 
  B_{\bf q}~(a_{\bf q}~a_{-\bf q}+a_{\bf q}^{\dagger}a_{-\bf
    q}^{\dagger})\, ,
\end{eqnarray}
with the coefficients
\begin{eqnarray}
  A_{\bf q} & = & SA+S(2J_{{\bf q}_1}'-J_{\bf q}'-J_{{\bf q}+{\bf
      q}_1}')+ S(2A^{xx}_{{\bf q}_1}-A^{yy}_{\bf q}-A^{zz}_{{\bf
      q}+{\bf q}_1})\label{Aq}\\ 
  B_{\bf q} & = & -SA+S(J_{{\bf q}+{\bf q}_1}'-J_{\bf q}')+ S(A^{zz}_{{\bf
  q}+{\bf q}_1}-A^{yy}_{\bf q})\label{Bq}\ .
\end{eqnarray}
Due to the large planar anisotropy $A$, for the experiments explained below, it
is suffucient to study only fluctuations within the hexagonal plane. The full
expression for arbitrary wave vectors will be given in \cite{pich96}. The wave
vector ${\bf q}_1=2\pi/\sqrt 3(0,1,0)$ (corresponding to $(\frac{1}{2},0,0)$ in
reciprocal lattice units (r. l. u.)) describes the antiferromagnetic modulation
of the ground state. After diagonalizing this Hamiltonian via a Bogoliubov
transformation we obtain the dispersion relation
\begin{eqnarray}
  E_{\bf q} & = & \sqrt{A_{\bf q}^2-B_{\bf q}^2}\nonumber \\
  & = & 2S \sqrt{(J_{{\bf q}_1}'-J_{\bf q}'+A^{xx}_{{\bf
  q}_1}-A^{yy}_{\bf q})(A+J_{{\bf q}_1}'- J_{{\bf q}+{\bf q}_1}'+A^{xx}_{{\bf
  q}_1}- A^{zz}_{{\bf q}+{\bf q}_1})}\, .
\label{dispequat}
\end{eqnarray}
Equation~(\ref{dispequat}) holds for the crystallographic Brillouin zone
(hexagon)\@. In the smaller magnetic Brillouin zone (rectangular, see
Fig.~(\ref{BrillFig})) there are two modes which have the form
\begin{mathletters}
\begin{eqnarray}
E^{(1)}_{\bf q} &=& \sqrt{A_{\bf q}^2-B_{\bf q}^2}\\
E^{(2)}_{\bf q} &=& \sqrt{A_{{\bf q}+{\bf q}_1}^2-B_{{\bf q}+{\bf q}_1}^2}\ .
\end{eqnarray}
\end{mathletters}

Stability of the ground state requires that, for all wave vectors in the
Brillouin zone, $A_{\bf q} > |B_{\bf q}|$, i.e. 
\begin{equation}
J_{{\bf q}_1}'-J_{\bf q}' > A^{xx}_{{\bf q}_1}-A^{yy}_{\bf q}\, .
\end{equation}
This condition gives an upper boundary for the exchange energy for
${\bf q} = 0$ and a lower bound for ${\bf q}={\bf q}_0 =
4\pi/3(1,0,0)$ (or ${\bf q} = (-\frac{1}{3},\frac{2}{3},0)$ in r.l.u.)
\begin{equation}
  A^{xx}_{{\bf q}_1}-A^{xx}_{{\bf q}_0} < J'< (A^{xx}_{{\bf
      q}_1}-A^{yy}_0 )/8 \, .
\end{equation}
Note that the allowed range for the exchange energy depends (due to the
restriction to $q_z = 0$) explicitly neither on the ferromagnetic
exchange $J$ nor on the anisotropy energy $A$\@. Using the in-plane
lattice constant $a=6.21 $\AA\ and the experimentally determined
Land\'e factor $g=2.25$ \cite{STEI91} of CsNiF$_3$, the stability
range for the exchange energy can be calculated to be\cite{footnote}

\begin{equation}
  -92{\rm mK} < \frac{J'}{k_B} < -3 {\rm mK}\, .
\label{stab}
\end{equation}

\subsection{Scattering amplitudes}

The inelastic magnetic scattering cross section is proportional to
\cite{lovesey}
\begin{equation}
  {\partial^2 \sigma\over \partial \Omega\partial\omega} \propto
  \sum_{\alpha\beta} \left|F({\bf Q})\right|^2\left(\delta^{\alpha\beta} - {Q_\alpha
      Q_\beta\over Q^2} \right) \int_{-\infty}^{\infty}{dt\over 2\pi}
      e^{-i\omega t} \left\langle S^\alpha_{\bf q}(t) 
        S^\beta_{-\bf q}\right\rangle \, .
\end{equation}
Here ${\bf Q}$ denotes momentum transfer (scattering vector), $F({\bf
  Q})$ is the magnetic form factor
and ${\bf q}$ the wave vector to the nearest reciprocal lattice vector
or position in the Brillouin zone $\tau$ (${\bf Q} = {\bf \tau} + {\bf
  q}$)\@. In linear spin wave theory the spin-spin correlation
functions can be evaluated with the Fourier transformed
Eqs.~(\ref{trafo1},\ref{trafo2},\ref{trafo3}) and the Bogoliubov
transformation. The cross section takes the form:

\begin{eqnarray}
  {\partial^2 \sigma\over \partial \Omega\partial\omega} & \propto &
\left|F({\bf Q})\right|^2\left[ 
 \left(1-{Q_y^2\over Q^2}\right){(A_{\bf q}-B_{\bf q}) 
      \over E_{\bf q}^{(1)}}\left\{(1+n_{\bf q})\delta\left(\omega-
        E_{\bf q}^{(1)}\right)+n_{\bf q}\delta\left(\omega+
        E_{\bf q}^{(1)}\right)\right\}+\right. \nonumber \\
  & & \nonumber \\
  & &\left.\left(1-{Q_z^2\over Q^2}\right){(A_{{\bf q}+{\bf q}_1}+
      B_{{\bf q}+{\bf q}_1}) 
      \over E_{\bf q}^{(2)}}\left\{(1+n_{\bf q})\delta\left(\omega-
        E_{\bf q}^{(2)}\right)+n_{\bf q}\delta\left(\omega+
        E_{\bf q}^{(2)}\right)\right\}\right]\, .
\label{scattering}
\end{eqnarray}
For neutrons, only spin fluctuations perpendicular to the momentum
transfer {\bf Q} are detectable, meaning modes with magnetization
vector parallel to the momentum transfer are invisible. Note that the
first mode $ E_{\bf q}^{(1)}$ is only observable through the in-plane
fluctuations $\langle S^y S^y\rangle$ and the second mode $ E_{\bf
  q}^{(2)}$ through the out-of-plane fluctuations $\langle S^z S^z\rangle$\@.
Thus, the first mode ($ E_{\bf q}^{(1)}$) will be called in-plane mode
and the second mode ($ E_{\bf q}^{(2)}$) out-of-plane mode. Due to the
strong planar anisotropy ($A$) the in-plane fluctuations are more
pronounced than the out-of-plane fluctuations, which can be seen by
inserting Eqs.~(\ref{Aq}) and (\ref{Bq}) for $A_{\bf q}$ and $B_{\bf
  q}$\@. The ratio of the two prefactors is given by
\begin{equation}
  {A_{\bf q}-B_{\bf q}\over A_{{\bf q}+{\bf q}_1}+B_{{\bf q}+{\bf
        q}_1}} = {A+J_{{\bf q}_1}'-J_{{\bf q}+{\bf q}_1}+A^{xx}_{{\bf
        q}_1}-A^{yy}_{{\bf q}+{\bf q}_1}\over J_{{\bf q}_1}'-J_{{\bf
        q}+{\bf q}_1}+A^{xx}_{{\bf q}_1}-A^{zz}_{{\bf q}+{\bf q}_1}}\,
  .
\end{equation}
This leads to a very small neutron scattering cross section of the
out-of-plane mode.

\section{Experiment}
The measurements were carried out using the cold source triple axis
spectrometer V2 at BENSC (Hahn-Meitner-Institut)\@.  Pyrolytic
graphite (PG) crystals were used as monochromator and analyzer.  The
higher-order wavelength contributions were suppressed by using a
liquid-N$_2$ cooled Be-filter. The crystal had a volume of about 1.5
cm$^3$, and was mounted with the ($a^*,c^*$) plane in the scattering
plane.  A series of constant-Q scans at positions $(Q_a,0,0)$ and
$(Q_a,0,2)$ were performed at $T=1.5$~K.  The final energy at the
$Q_c=0$ positions ($(Q_a,0,0)$ scans) was fixed to be $E_f= 2.98$ meV
(collimation: neutron guide (NG) -40'-40'-40')\@. The collimation of
the neutron guide for the used values of $k_i$ is approximately 60'.
At the $Q_c=2$ positions ($(Q_a,0,2)$ scans) the final energy was
increased to $E_f = 4.66$ meV (collimation: NG-20'-20'-20')\@.  The
capital letters $Q_{a,b,c}$ denote absolute Q values, while
$q_{a,b,c}$ represents the relative distance to the center of the
Brillouin zone. At all $Q_c=0$ positions only one excitation peak was
observable. The data at ${\bf Q} = (0.8,0,0)$ ({\bf q}~=~(0.2,0,0)) is
shown as a representative example in Fig.~\ref{rawdat0}\@.  The
profile of the incoherent background (centered at E~=~0) and the
excitation signal were fitted by Gauss-peaks. The line widths are
consistent with the expected instrument resolution. As discussed in
the previous section, the in-plane mode has a much larger scattering
amplitude ($\sim 6$ times) than the out-of-plane mode. Nevertheless,
the in-plane mode from domain A cannot be detected, because ${\bf Q}$
is parallel to ${\bf y}$ (see fig.~\ref{Spinstruct0})\@. Thus, only
the in-plane modes from domains B and C should be visible.

This was probed by a separate measurement in a horizontal magnetic
field. In zero field all three magnetic domains of the crystal have
approximately the same size. The relative volume parts of the
different domains can be varied by applying an external magnetic field
perpendicular to the c-axis \cite{STEI74}\@. A horizontal field
parallel to $a^*$ stabilizes domain A by the possibility of a slight
spin canting.  This is shown by the increasing intensity of the
(0.5,0,0) Bragg reflection, when increasing the magnetic field (insert
of Fig.~\ref{magdat})\@. At a field of about 700~Gauss only domain A
remains.  Higher fields give rise to an increased spin canting,
leading to a paramagnetic phase above 3000~Gauss.  Figure~\ref{magdat}
shows the spin wave excitation at {\bf Q}~=~(0.7,0,0) ({\bf
  q}~=~(0.3,0,0)) for three different small magnetic fields. The
narrow windows of the used cryomagnet limited the final energy to the
value $E_f = 4.06$~meV ($ k_f = 1.4$\AA$^{-1}$)
(collimation:~NG-40'-40'-40')\@.  Thus, the resolution was lower in
this experiment compared to the zero field measurements performed with
$E_f = 2.98$~meV ($ k_f = 1.2 \mbox{\AA}^{-1}$)\@.  Obviously, the
increase in the magnetic field reduces the intensity of the excitation
, which confirms that the excitations at ${\bf Q}=
(Q_a,0,0)$ arise from the magnetic domains B and C.

For measuring the in-plane-mode in domain A, one has to use a momentum
transfer {\bf Q} not parallel to the $a^*$-axis ($y$-direction)\@.
This was done by choosing ${\bf Q}=(Q_a,0,2)$ (The size of the
magnetic Brillouin zone in $c^*$ direction is twice the size of the
crystalline Brillouin zone)\@.  Unfortunately, the high Q values
restricted $E_f$ to large values ($E_f = 4.66$~meV;~$ k_f = 1.5
\mbox{\AA}^{-1}$). This caused a coarse resolution compared to the
measurement at ${\bf Q}=(Q_a,0,0)$ even with a better collimation
(NG-20'20'20')\@.  The measurement at ${\bf Q}=(0.8,0,2)$ ({\bf
  q}~=~(0.2,0,0)) is shown in (Fig.~\ref{rawdat2})\@.  At first
glance, there seems to be just one excitation at about 0.15 meV.
However, knowing the existence of an excitation at 0.118 meV from the
measurement at ${\bf Q} = (0.8,0,0)$ it is possible to fit a second
excitation at 0.175(5) meV.  The fit includes: two Gaussian peaks with
fixed energy ($\pm$0.118 meV), one Gaussian peak for the incoherent
background (E~=~0~meV) and one Gaussian peak for the second
excitation. The widths of the different Gaussians were fitted
independently. As for the measurements at $(Q_a,0,0)$, the line widths
are caused by instrument resolution.  All other measurements at
($Q_a$,0,2) were treated in the same way, except the measurement at
(0.6,0,2) where the widths of all Gaussians were set equal.

\section{Discussion and Conclusion}

The calculation of the dispersion relation was performed for the spin
configuration of domain A. For comparison of the measured data with
the theory, it is convenient to transform the signals from domains B
and C to equivalent points in domain A. This can be done simply by
rotations of the reciprocal lattice through $\pm 60^\circ$, which
change the measured $Q$ positions from $(q_a,0,0)$ to $(0,q_b,0)$\@.
The data obtained from the measurements at $(Q_a,0,2)$ belong already
to domain A.

All measured data points of the dispersion relation are plotted in 
Fig.~(\ref{dispfig})\@. The theoretical dispersion relation derived in
section II (Eq.~(\ref{dispequat})) was fitted to these experimental
data.  The fit included just two free parameters: the value of the
interchain exchange interaction $J'$ and the easy-plane-anisotropy
$A$\@. Good agreement between theory and experiment can be obtained with
the values:

\begin{eqnarray}
J'/k_B & = & -0.0247(12)K \nonumber\\
A /k_B & = & 3.3(1)K\, . \nonumber
\end{eqnarray}

The determined value of $J'$ is consistent with the condition for the
stability of the ground state (Eq.~(\ref{stab}))\@. It turns out that CsNiF$_3$
is far away from the transitions mentioned in section I, and thus neglecting
higher order terms in the Holstein-Primakoff transformation is expected to be a
a reliable low-temperature approximation. The value for the
easy-plane-anisotropy $A$ of 3.3~K is lower than an earlier value
($A_{1D}/k_B=4.5$~K), determined from neutron scattering experiments
using a linear spin wave theory above T$_N$\@.  The actual difference
is even larger, because the new value represents the pure crystal
field anisotropy, while in the old measurements the effect of the
intrachain dipolar interaction was not separated from the single site
anisotropy.  $A_{1D}$ is an effective anisotropy. To compare these two
values one has to calculate the dipolar anisotropy $D$ in isolated
spin chains, which leads to $D \approx -0.64$~K. The calculation of D
is possible by assuming a strictly ferromagnetic ordering of the spins
along the magnetic chain. The good convergence of dipolar sums in one
dimension, causes this value to be reached even for short-range
ordered chains.  The easy-axis-type dipolar anisotropy $D$ has to be
added to the easy-plane pure crystal field anisotropy $A$ to give the
old value of $A_{1D}~=~D~+~A~=~4.5$~K.  Thus, the value for $A$ from
the measurements at $T > T_N$ is $A/k_B = 5.1$~K.  A possible source
of this difference is the neglect of the dipolar inter-chain
interaction in the model used in the temperature range above T$_N$\@.
Maybe, an independent determination of $A$ by measuring the dispersion
relation along the c-direction in the long-range ordered
antiferromagnetic state ($T < T_N$) is necessary to solve this
problem.

The influence of each parameter of the Heisenberg--Hamiltonian
(\ref{hamilton})  on the dispersion relation is directly
visible at characteristic points of the Brillouin zone. Inserting
Eq.~(\ref{Jstrichq}) into Eq.~(\ref{dispequat}) gives the following
expressions for the excitation energies at the {\bf Q}--positions of
$\Gamma$, P, X and S:
\begin{eqnarray}
  E^{(1)}_{\Gamma} = E_{\Gamma} & = & 2S\sqrt{(-8J'+ A^{xx}_{P}
    -A^{yy}_{\Gamma})(A + A^{xx}_{P}-A^{zz}_{P})} \\ 
  E^{(2)}_{\Gamma} = E_{P} & = & 2S\sqrt{(A^{xx}_{P}-A^{yy}_{P})(A-8J' +
  A^{xx}_{P}-A^{zz}_{P})} \\ 
  E_{S} & = & 2S\sqrt{(-4J'+A^{xx}_{P}-A^{yy}_{S})(A-4J' +
    A^{xx}_{P}-A^{zz}_{S})} \\ 
  E_{X} & = & 2S\sqrt{(A^{xx}_{P}-A^{yy}_{X})(A+A^{xx}_{P}-A^{zz}_{X})}
\end{eqnarray}
The values of the gap at the point $\Gamma$ are determined by the
single ion anisotropy and the dipolar interaction. The in-plane mode
E$^{(1)}_\Gamma$ exhibits a gap even without dipolar terms, but the
gap of the out-of-plane mode E$^{(2)}_\Gamma$ vanishes in case of no
dipolar interaction.

Two features of the dispersion relation are very unusual, and
demonstrate the strong influence of the dipolar interaction in
CsNiF$_3$: the position of the minimum of the dispersion relation and
the jump of the dispersion relation at the Brillouin zone center.
Contrary to common spin wave dispersion relations the energy minimum
is not found at the magnetic zone center, but near the magnetic
Brillouin zone boundary.  This is due to the strong anisotropy of the
dipolar interaction and the competition of the dipolar and exchange
interaction.

The jump in the dispersion relation at the zone center $\Gamma$ as
shown in Fig.~(\ref{dispfig}) can be viewed at in a similar way as
the well known splitting of the longitudinal and transverse optical
phonon modes in polar crystals. The LO--TO--splitting is caused by
electric long-range dipolar interactions. This is used, for instance,
for the derivation of the Lyddane--Sachs--Teller relation in solid
state physics textbooks (e.g. \cite{ASHC})\@.  The splitting is a result
of the semi-convergence of dipole sums in homogeneously polarized
systems. This gives rise to a depolarization field for longitudinal
phonons with long wavelength, but not for transverse modes.

Similar arguments are valid for spin waves in CsNiF$_3$\@.  Here, not
only the large influence of the dipolar interaction is important, but
also the planar anisotropy, leading to a linear polarized dynamic
magnetization in the out-of-plane and in-plane modes.  This allows to
describe the in-plane mode in analogy to phonons as ``longitudinal''
or ``transverse'' spin wave, depending on the angle between dynamic
magnetization (${\bf q}_a$-direction) and propagation direction of the
wave (Fig.~\ref{MagPat}). In common notation both are transverse spin
fluctuations in respect to the spin orientation. As in the case of
phonons, the ``longitudinal'' spin waves have the highest energy (wave
propagation along ${\bf q}_a$)\@. In the observed plane of the Q-space
($q_c = 0$) the out-of-plane mode is ``transverse'' for all spin wave
propagation directions. Thus, the dispersion relation of this mode
exhibits no jump at $\Gamma$\@.

In summary it is shown, that the description of the spin system of
CsNiF$_3$ including long-range dipolar interactions gives a convincing
explanation for the unusual antiferromagnetic structure and spin
dynamics. Especially, peculiar features of the spin wave dispersion
relation can only be explained by a strong influence of long-range
dipolar interactions in CsNiF$_3$\@. The jump of the spin wave
dispersion relation at the zone center has been so far observed only
in very few ferromagnets \cite{JEMA}\@. CsNiF$_3$ is the first
antiferromagnet exhibiting this feature. This is caused by special
properties of the one dimensional spin system CsNiF$_3$\@. Firstly,
the strong 1D ferromagnetic order leads to a dipolar interchain
interaction in the same order as the weak exchange interaction $J'$\@.
Secondly, the planar anisotropy enforces a special spin dynamic, which
gives rise to a dynamical magnetization pattern in long-wavelength
spin waves. This demonstrated once again that low dimensional magnets
are very suitable model systems to study a wide range of fundamental
magnetic properties.

\acknowledgments This work has been supported by the German Federal
Ministry for Research and Technology (BMBF) under the contract number
03-SC4TUM

\figure{\label{Spinstruct0}The three different magnetic domain types
  in CsNiF$_3$\@. \hfill}

\figure{\label{BrillFig}The crystallographic
  (\rule[0.5ex]{3.0em}{0.5pt}) and magnetic ($-----$) Brillouin zones
  of CsNiF$_3$ in the ($a^*, b^*$)-plane. The solid circles
  ($\bullet$) indicate positions of nuclear Bragg reflections while
  the open circles ($\circ$) mark the positions of the magnetic Bragg
  reflections.}

\figure{\label{rawdat0} Measured data and fit at Q = (0.8,0,0);
  ($-----$) magnetic excitation and incoherent background;
  (\rule[0.5ex]{3.0em}{0.5pt}) sum signal plus  background.}

\figure{\label{magdat} Field dependence of signal at Q = (0.7,0,0).
  The solid lines are guides to the eye.  The Insert shows field
  dependent intensity of the magnetic reflection at (0.5,0,0)\@. }

\figure{\label{rawdat2} Measured data and fit at Q = (0.8,0,2);
  ($-----$) magnetic excitation of domains (B),(C) with fixed energy
  determined by the measurements at (0.8,0,0);
  $(\cdot\cdot\cdot\cdot\cdot\cdot)$ magnetic excitation of domain (A)
  and incoherent background; (\rule[0.5ex]{3.0em}{0.5pt}) sum signal
  plus background.}

\figure{\label{dispfig} The dispersion relation fitted to the measured
  data. The numbers at the lower abscissa denotes the {\bf q}-position
  along the $a^*$ and $b^*$ axes in the first Brillouin zone.  The
  letters at the upper abscissa correspond to special points in the
  Brillouin zone (Fig. (2))\@. The in-plane mode $E_q^{(1)}$ and
  out-of-plane mode $E_q^{(2)}$ are shown by the solid and dashed
  lines, respectively. A striking feature is the jump of the
  in-plane mode at $\Gamma$.}

\figure{\label{MagPat} Schematical representation of the spin wave
  induced dynamical in-plane magnetization ($\Rightarrow$) for long
  wavelengths. Two cases are shown: wave vector {\bf k} parallel (a) and
  perpendicular (b) to the ordered moments. The dipole energy differs
  between both cases, leading to the jump of the dispersion relation at
  $\Gamma$.}

\newpage

\epsfysize = 17cm
\rotate[l]{\epsffile{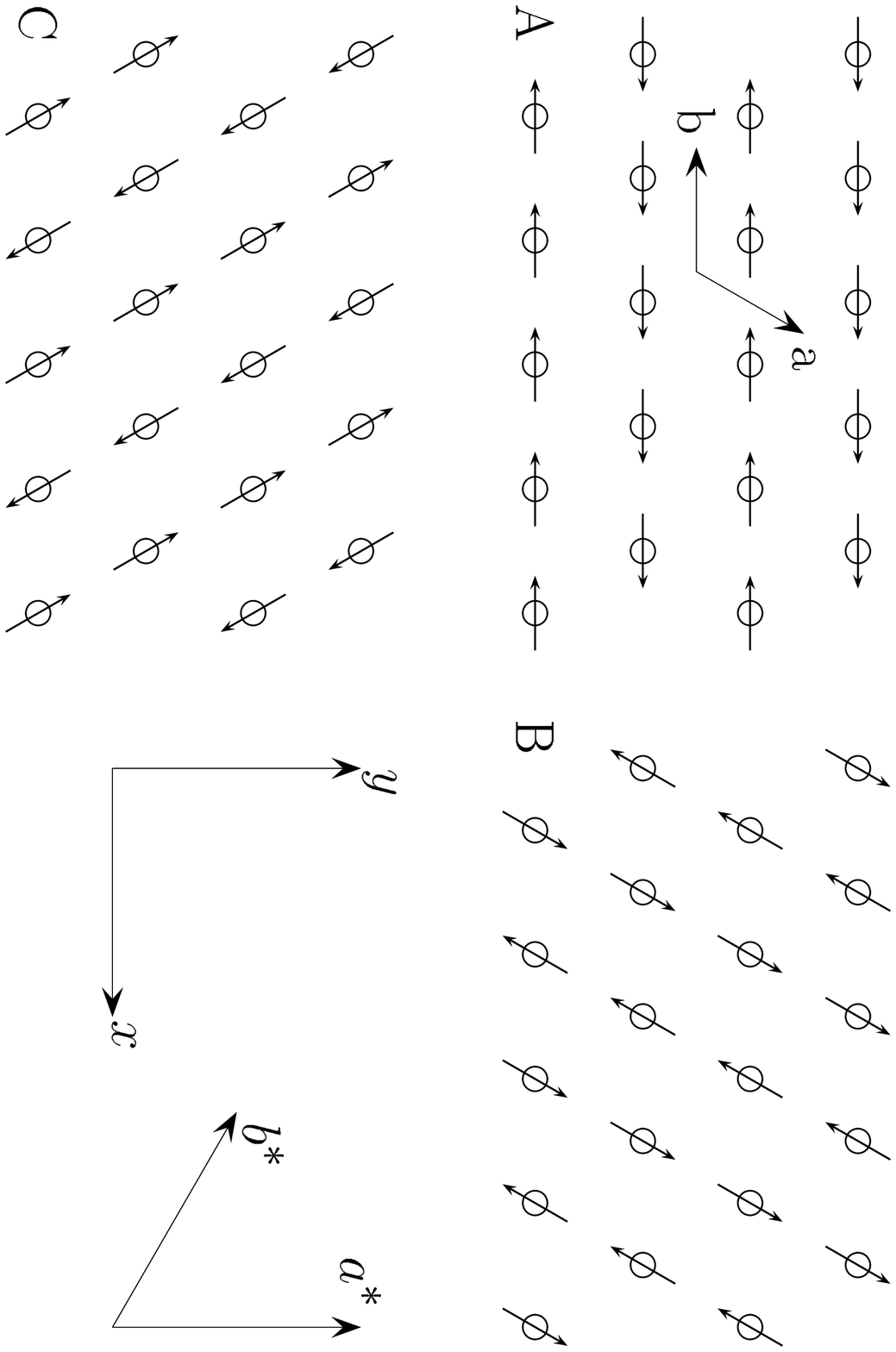}}
Fig. 1

\epsfysize = 17cm
\rotate[l]{\epsffile{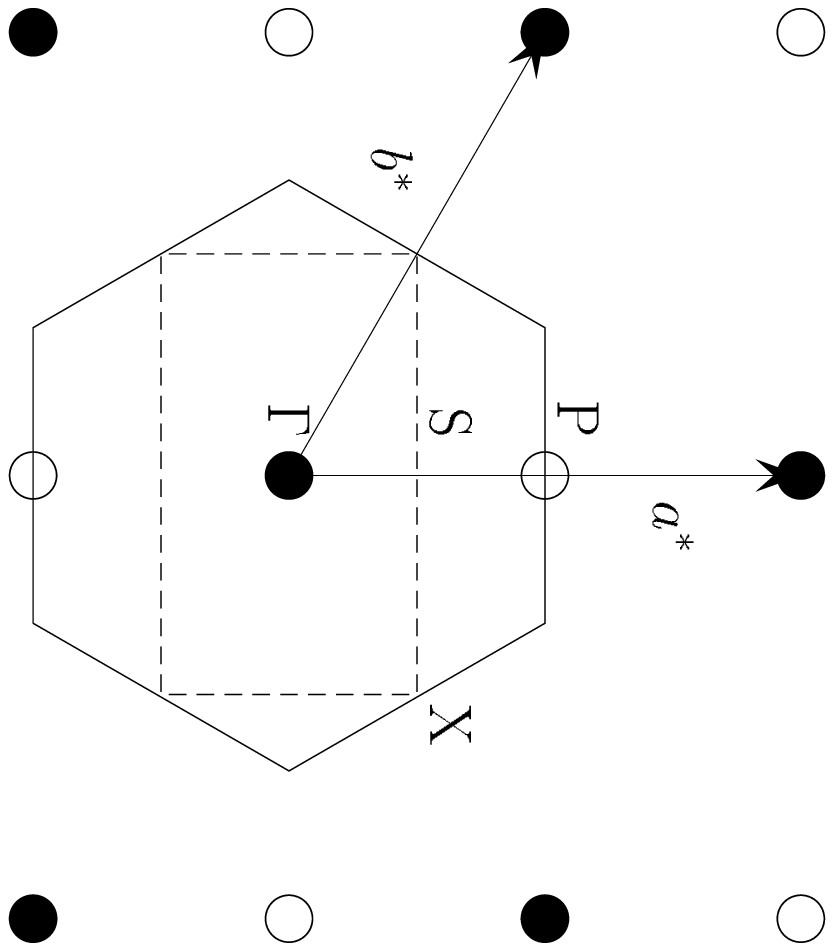}}
Fig. 2
\vskip 1cm
\epsfysize = 8cm
\epsffile{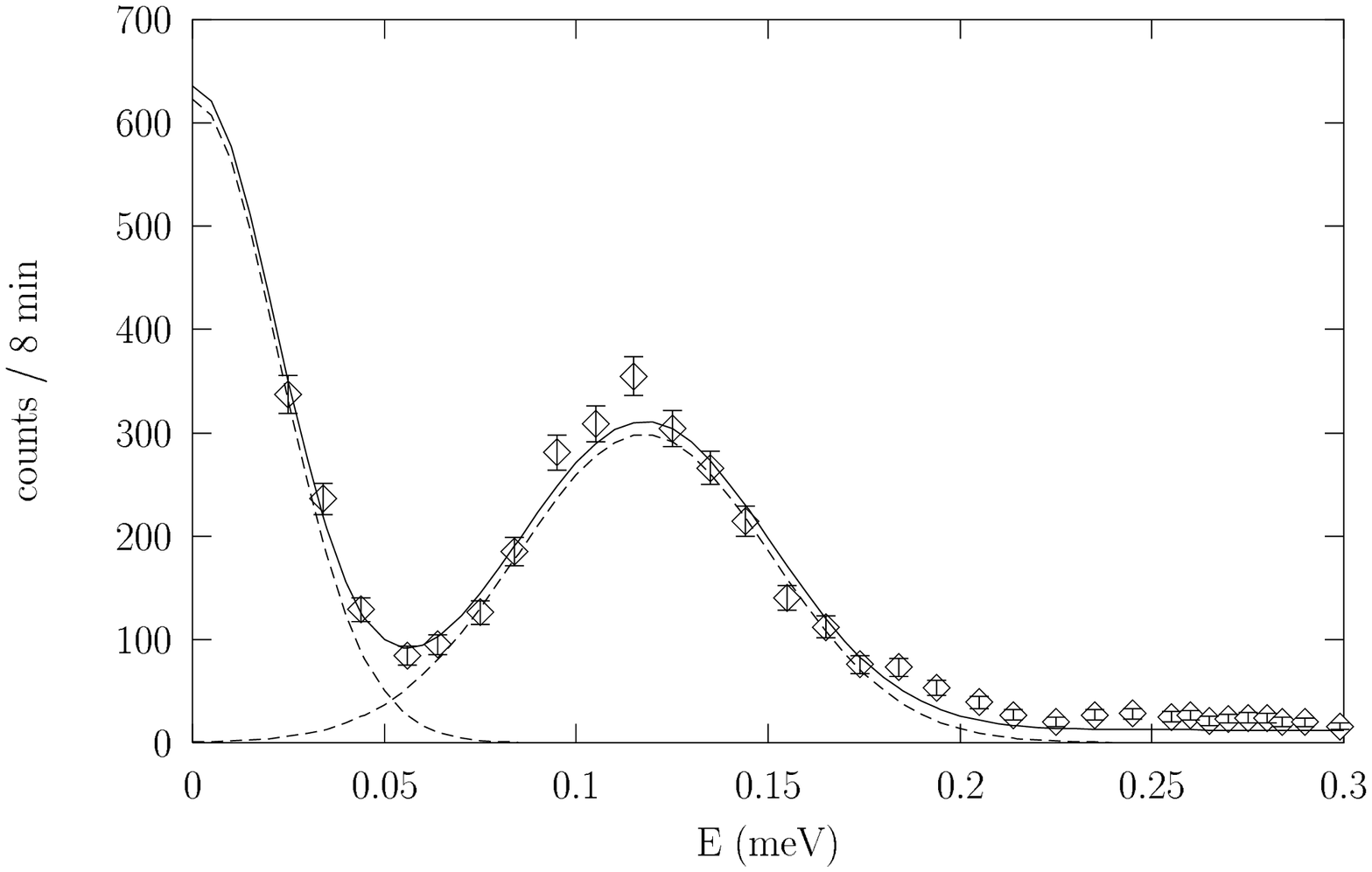}

Fig. 3

\epsfysize = 9cm
\epsffile{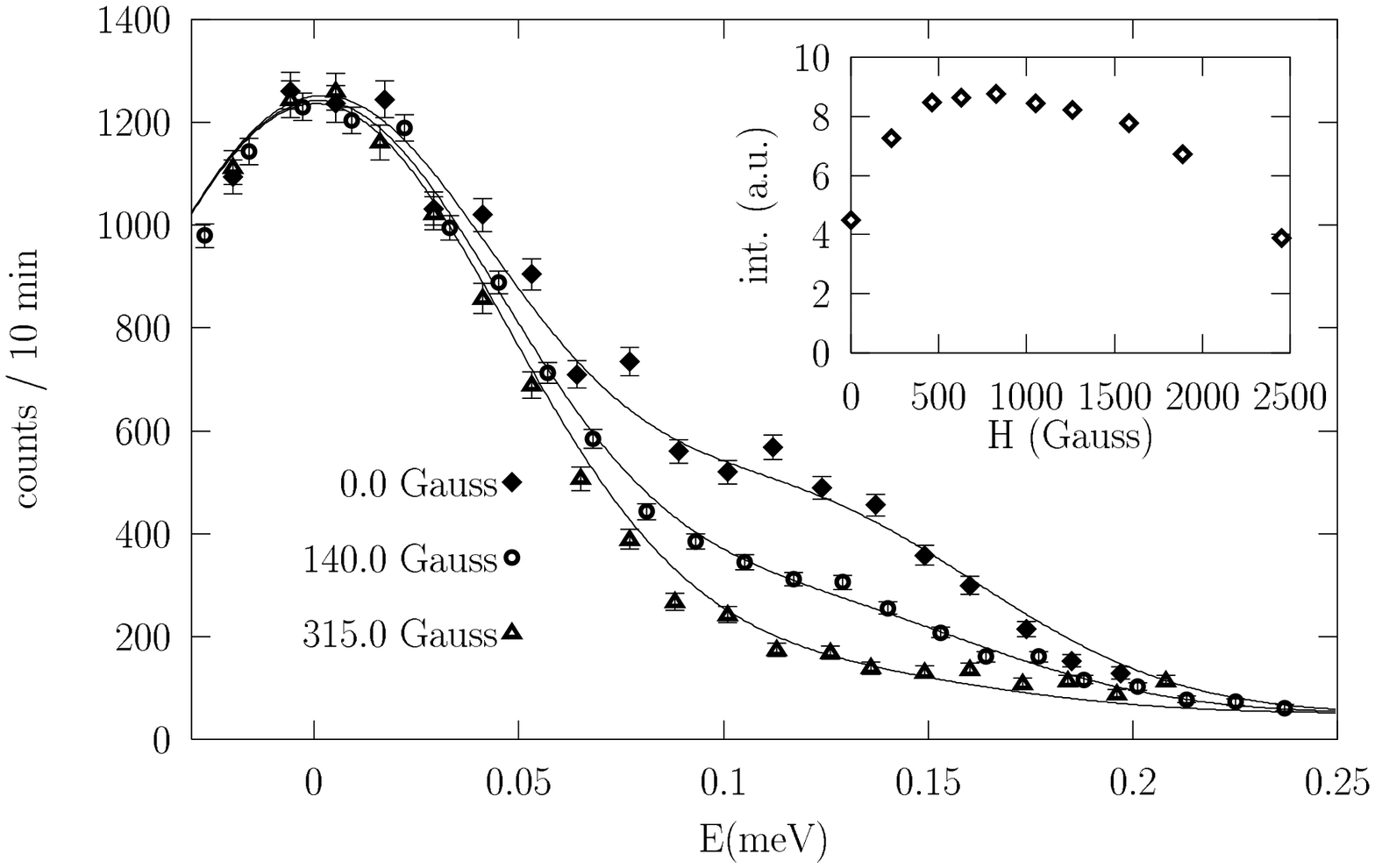}
Fig. 4
\vskip 2cm
\epsfysize = 9cm
\epsffile{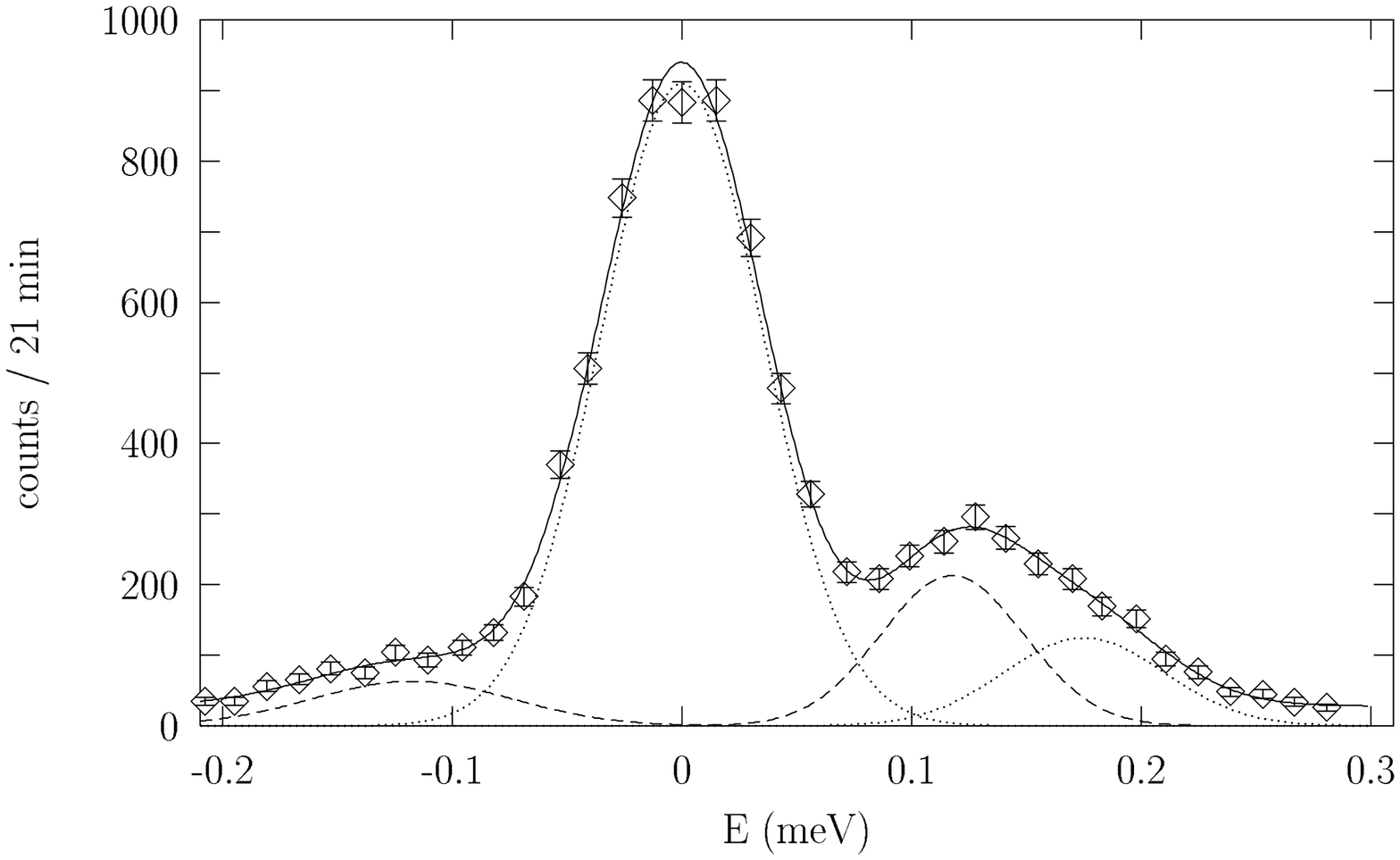}
Fig. 5

\epsfysize = 10cm
\epsffile{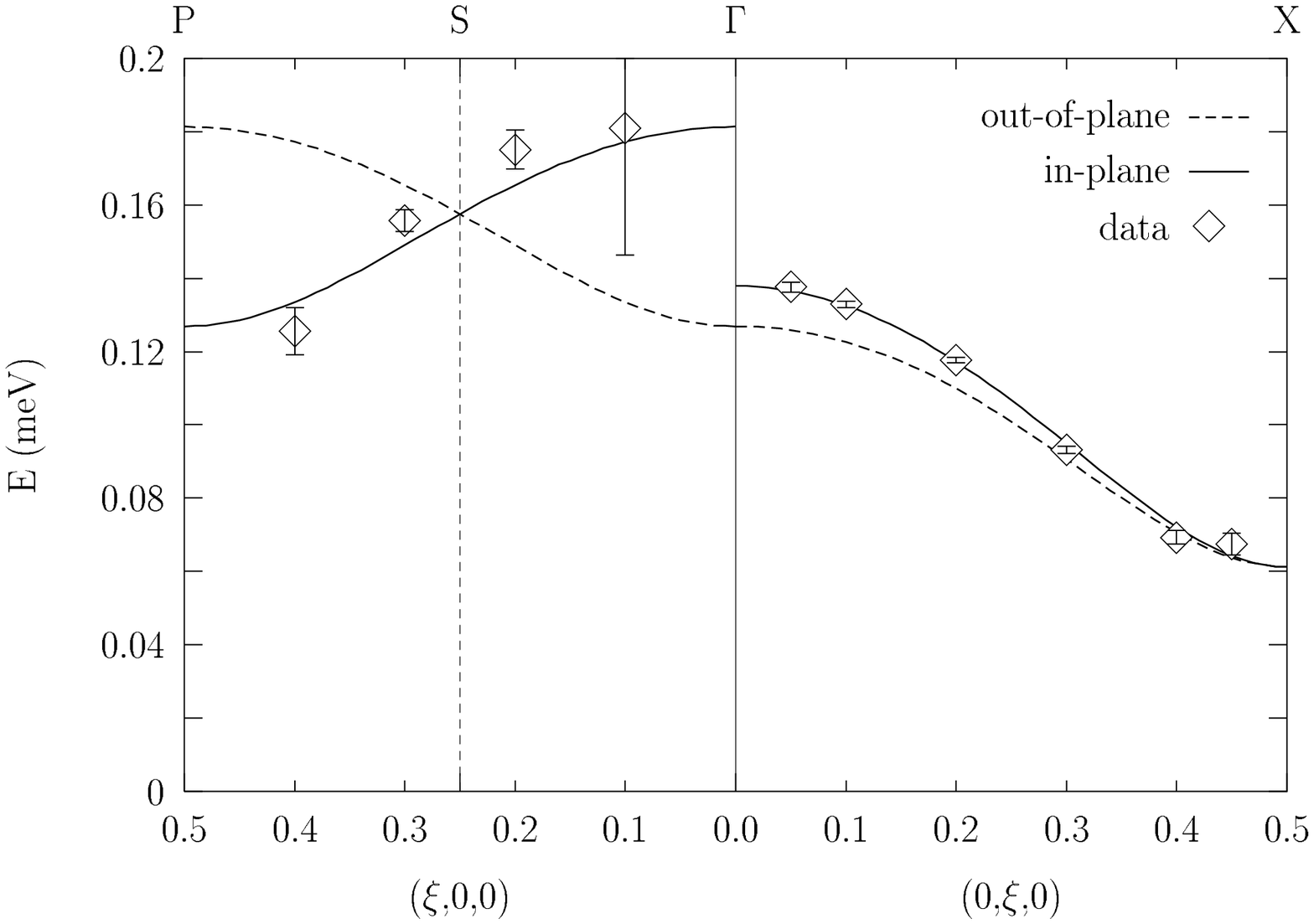}

Fig. 6
\vskip 2cm
\epsffile{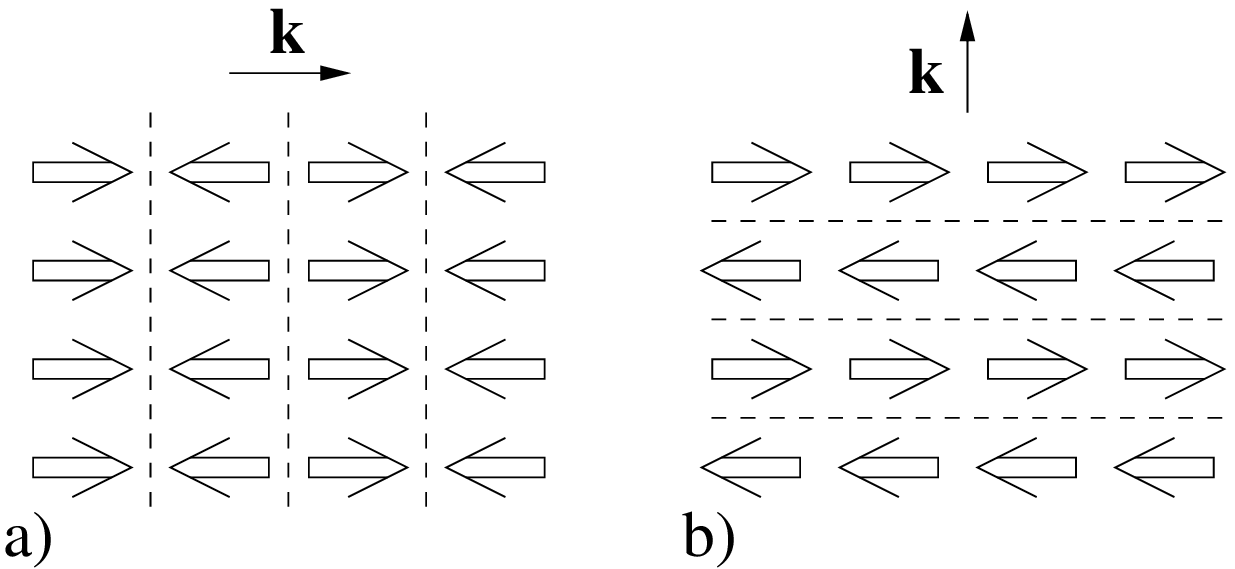}

Fig. 7

\end{document}